\documentclass[letterpaper,10pt]{article}

\usepackage[utf8]{inputenc}
\usepackage{graphicx}
\usepackage{psfrag}

\newcommand{\dd}{\textrm{d}}

\newcommand{\chypergf}{{}_{1}F_{1}}

\title{New conditionally exactly solvable inverse power law potentials}

\author{A.\ L\'opez-Ortega \\
Departamento de F\'{\i}sica.\\ 
Escuela Superior de F\'{\i}sica y Matem\'aticas. \\
Instituto Polit\'ecnico Nacional. \\
Unidad Profesional Adolfo L\'opez Mateos. Edificio 9. \\
M\'exico, D.\ F., M\'exico. \\
C.\ P.\ 07738 \\
email: alopezo@ipn.mx }

\begin{document}

%

\maketitle

\begin{abstract}

We give two conditionally exactly solvable inverse power law potentials whose linearly independent solutions include a sum of two confluent hypergeometric functions. We notice that they are partner potentials and multiplicative shape invariant. The method used to find the solutions works with the two Schrodinger equations of the partner potentials. Furthermore we study some of the properties of these potentials.



\end{abstract}




\section{Introduction}
\label{s: Introduction}

In physics the search and study of exactly solvable systems have a long history. In non relativistic quantum mechanics a problem thoroughly investigated is to find potentials for which we can solve exactly their Schr\"odinger equations in terms of special functions, since they can be used as a basis in the analysis of more realistic problems and their solutions can be exploited to test numerical calculations. Several methods have been developed to obtain exact solutions to the Schr\"odinger equation \cite{Bhattacharjie}--\cite{Infeld-Hull}, here we mention the methods based on the point canonical transformations \cite{Bhattacharjie}, on the Darboux transformations \cite{Darboux}, \cite{Bagrov}, and on the supersymmetric quantum mechanics \cite{Witten-susy}--\cite{Bagchi-book} (see also the  factorization method \cite{Schrodinger}--\cite{Infeld-Hull} related to the supersymmetric quantum mechanics). 

At present time we know many potentials for which  we can solve exactly their Schr\"odinger equations \cite{Dutt-ajp-1988}--\cite{Bagchi-book}, \cite{Khare-scattering}--\cite{Levai-search}. Furthermore, there are potentials for which we can find their solutions in terms of special functions only if the parameters satisfy some condition \cite{Souza Dutra}--\cite{Roychoudhury}. They are called conditionally exactly solvable potentials (CES potentials).  

In this work, our main objective is to give two potentials defined on the half line $x \in (0,\infty)$ such that each linearly independent solution of the Schr\"odinger equations includes a sum of two confluent hypergeometric functions. As far as we know these potentials are not previously studied \cite{Dutt-ajp-1988}--\cite{Bagchi-book}, \cite{Khare-scattering}--\cite{Roychoudhury}. Here we notice that the method used in this paper works simultaneously with the two Schr\"odinger equations of the partner potentials. 

The potentials that we study in this paper are algebraic modifications of the Coulomb potential (see the expression (\ref{e: ces potentials})). These potentials may be useful in the study of physical systems for which we expect deviations from the Coulomb potential near $r=0$, but that are of Coulomb type as $r \to \infty$. It is convenient to notice that potentials with negative fractional powers have been used as models to study the quark-antiquark interaction \cite{Song quark} and sharp resonances \cite{Stillinger potential}. From the shape of the potentials (see Figure 2 below) we think that these can be useful models to study scattering and tunneling problems \cite{Flugge}. Furthermore, in an interval the shape of the potential that we call $V_-$ reminds us to the effective potentials that appear in the analysis of the propagation of the Dirac field in curved spacetimes, for example, in the Schwarzschild black hole \cite{Cho-Dirac-qnms}. 

We organize this paper as follows. In Sect.\ \ref{s: Method} we expound the potentials that we study through this work and  we solve the Schr\"odinger equations of these potentials. We also find some properties of these solutions. Finally in Sect.\ \ref{s: Discussion} we discuss other properties of these potentials.

\section{Solution method}
\label{s: Method}

Here we show that for $x \in (0,+\infty)$ and for the potentials
\begin{eqnarray} \label{e: ces potentials}
V_\pm (x, m) &=&  \frac{m^2}{x} \pm \frac{m}{2} \frac{1}{x^{3/2}}  \\
&=& W^2 \pm \frac{\dd W}{\dd x}, \nonumber
\end{eqnarray} 
we can solve the Schr\"odinger equations
\begin{eqnarray} \label{e: Schrodinger equation}
\frac{\dd^{2} Z_{\pm}}{\dd x^{2}} + \omega^{2}  Z_{\pm} = V_{\pm}  Z_{\pm} , 
\end{eqnarray}
in terms of confluent hypergeometric functions.\footnote{To simplify some of the expressions that follow, we denote the energy $E$ as $\omega^2 $.} For the potentials (\ref{e: ces potentials}) the superpotential $W$ is equal to \cite{Cooper:1994eh}--\cite{Bagchi-book}
\begin{equation} \label{e: ces superpotential}
 W (x) = - \frac{ m}{\sqrt{x}} .
\end{equation} 
Thus, in the standard language of the supersymmetric quantum mechanics, $V_+$ and $V_-$ are partner potentials \cite{Cooper:1994eh}--\cite{Bagchi-book}. 

The partner potentials (\ref{e: ces potentials}) are CES since the constants $m^2$ and $\pm m/2$ that multiply to the factors $1/x$ and $1/x^{3/2}$ satisfy the condition
\begin{equation} \label{e: constriction CES}
 -\frac{1}{4} m^2 + \left( \pm \frac{m}{2} \right)^2 = 0 .
\end{equation} 
As far as we see the partner potentials (\ref{e: ces potentials}) do not appear in Refs.\ \cite{Dutt-ajp-1988}--\cite{Bagchi-book}, \cite{Khare-scattering}--\cite{Levai-search} that enumerate the previously known exactly solvable potentials of the Schr\"odinger equation. Moreover the CES  potentials (\ref{e: ces potentials}) remind us to those of Refs.\ \cite{Bagchi-book}, \cite{Souza Dutra}
\begin{eqnarray} \label{e: Souza Dutra potential}
 \hat{V}_{1}(x) &=& \frac{K}{x} + \frac{L}{x^{1/2}} - \frac{3}{16 x^2} , \nonumber \\
 \hat{V}_{2}(x) &=& M x^{2/3} + \frac{N}{x^{2/3}} - \frac{5}{36 x^2} ,
\end{eqnarray} 
where $K$, $L$, $M$, $N$ are constants, since they include negative fractional powers of the variable $x$ in addition to negative integral powers of the same variable.

\begin{figure}[th]
\label{figure1}
\begin{center}
\includegraphics[scale=1,clip=true]{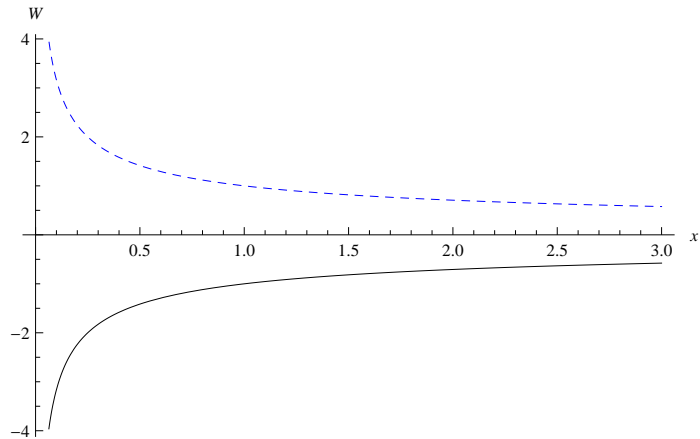}
\caption{Plots of the superpotential $W$ for $m=1$ (solid line) and for $m=-1$ (broken line).} 
\end{center}
\end{figure}

\begin{figure}[th]
\label{figure2}
\begin{center}
\includegraphics[scale=1,clip=true]{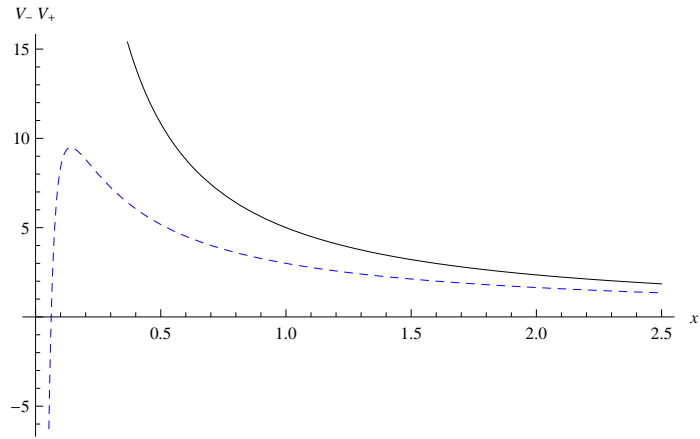}
\caption{Plots of the potential $V_+$ (solid line) and the potential $V_-$ (broken line) for $m=2$.} 
\end{center}
\end{figure}

In what follows we assume that $m > 0$. We notice that near $x=0$ and as $x \to +\infty$ the superpotential (\ref{e: ces superpotential}) and the CES potentials (\ref{e: ces potentials}) behave as
\begin{eqnarray} \label{e: limits potentials superpotential}
 \lim_{x \to 0^+} W &=& - \infty, \qquad \qquad  W_+ = \lim_{x \to + \infty} W = 0^-, \nonumber \\
 \lim_{x \to 0^+} V_+ &=& + \infty, \qquad \qquad  \lim_{x \to + \infty} V_{\pm} = 0^+ , \\
 \lim_{x \to 0^+} V_- &=& - \infty ,  \nonumber
\end{eqnarray} 
where $0^+$ ($0^-$) means that the quantity goes to zero taking positive (negative) values. We also point out that the CES potentials $V_\pm$ go to zero in the form $1/x$ as $x \to +\infty$, and they diverge as $1/x^{3/2}$ as $x \to 0^+$. The potential $V_+$ is strictly positive, whereas the potential $V_-$ crosses the $x$ axis at $x_0 = 1 / (4m^2)$. 

Since the derivatives of the partner potentials (\ref{e: ces potentials}) are equal to
\begin{equation}
 \frac{\dd V_\pm}{\dd x} = -\frac{m}{x^2}\left(m \pm \frac{3}{4 x^{1/2}} \right),
\end{equation} 
the potential $V_+$ is a decreasing function, and we get that the derivative of $V_-$ is equal to zero at $x_1 = 9 /(16 m^2)$. This critical point is a maximum. Notice that $x_1 > x_0$, that is, the coordinate of the maximum is larger than the point where the potential $V_-$ crosses the $x$ axis.  For $x \in (0,+\infty)$ the superpotential $W$ is strictly negative and it is an increasing function. For  $W$ and $V_\pm$ we illustrate these facts in Figs.\ 1, 2. 

For $\omega \neq 0$  a straightforward calculation shows that  the Schr\"odinger equations (\ref{e: Schrodinger equation}) for partner potentials  transform into
\begin{eqnarray}
 \left( \frac{\dd }{\dd x } \mp W \right) \frac{1}{i \omega} \left( \frac{\dd }{\dd x } \pm W \right) Z_\mp = i \omega Z_\mp .
\end{eqnarray}
From these equations we get that the functions $Z_\pm$ must be solutions to the coupled system of first order differential equations
\begin{eqnarray} \label{e: Z coupled}
\left( \frac{\dd }{\dd x } \pm  W \right) Z_\mp = i \omega Z_\pm, 
\end{eqnarray}
and defining the functions $\tilde{R}_1$ and $\tilde{R}_2$ by 
\begin{equation} \label{e: R tilde definition}
 Z_\pm = e^{-i \pi /4} (\tilde{R}_1 \pm i \tilde{R}_2),
\end{equation} 
we find that the Schr\"odinger equations (\ref{e: Schrodinger equation}) of partner potentials simplify to the system of coupled differential equations for the functions $\tilde{R}_1$ and $\tilde{R}_2$  
\begin{eqnarray} \label{e: coupled tilde R}
\frac{\dd \tilde{R}_{2} }{\dd x} + i\omega \tilde{R}_{2} &=& -i W \tilde{R}_{1} ,   \nonumber \\
  \frac{\dd \tilde{R}_{1} }{\dd x} - i\omega \tilde{R}_{1} &=& i W \tilde{R}_{2}.
\end{eqnarray}

Thus to solve the Schr\"odinger equations of the partner potentials (\ref{e: ces potentials}) we make the change of variable 
\begin{equation} \label{e: y definition}
 y = -2 i \omega x ,
\end{equation} 
to obtain that Eqs.\ (\ref{e: coupled tilde R}) take the form
\begin{eqnarray} \label{e: coupled tilde R variable y}
 \frac{\dd \tilde{R}_{2} }{\dd y} - \frac{1}{2} \tilde{R}_{2}  &=& - \frac{ m}{(2  \omega)^{1/2} i^{1/2} y^{1/2}} \tilde{R}_{1} ,  \nonumber \\
\frac{\dd \tilde{R}_{1} }{\dd y} + \frac{1}{2} \tilde{R}_{1}  &=& \frac{ m}{(2  \omega)^{1/2} i^{1/2} y^{1/2}} \tilde{R}_{2} .
\end{eqnarray}
From this coupled system we get that the functions $\tilde{R}_1$ and $\tilde{R}_2$ satisfy the decoupled equations 
\begin{equation} \label{e: r tilde differential equations}
 \frac{\dd^2 \tilde{R}_j}{\dd y^2} + \frac{1}{2 y}\frac{\dd \tilde{R}_j}{\dd y} +\left[ \left( \frac{\epsilon}{4} - \frac{m^2 i}{2 \omega} \right)\frac{1}{y} - \frac{1}{4} \right] \tilde{R}_j = 0,
\end{equation}
where $j=1,2,$ and $\epsilon = 1$ ($\epsilon = -1$) for $\tilde{R}_1$ ($\tilde{R}_2$). To find the solutions to the previous equations we make the ansatz
\begin{equation} \label{e: R tilde exponential}
 \tilde{R}_j = e^{-y/2} \hat{R}_j,
\end{equation} 
to get that the functions $\hat{R}_j$ must be solutions to 
\begin{equation} \label{e: R radial y coordinate}
 y \frac{\dd^2 \hat{R}_j}{\dd y^2} + \left( \frac{1}{2} - y \right) \frac{\dd \hat{R}_j}{\dd y} - \left( \frac{1 - \epsilon}{4} + \frac{i m^ 2}{2 \omega} \right) \hat{R}_j = 0.
\end{equation} 

We notice that the previous equations are of confluent hypergeometric type \cite{Abramowitz-book}--\cite{NIST-book}
\begin{equation} \label{e: confluent hypergeometric equation}
 y \frac{\dd^2 F }{\dd y^2} + (b -y)\frac{\dd F}{\dd y} - a F = 0,
\end{equation}
with the quantities $a_j$ and $b_j$ equal to
\begin{eqnarray} \label{e: values parameters a b}
 a_1 &=& a_2 -\frac{1}{2} = \frac{i m^2}{2 \omega} , \nonumber \\
 b_1 &=& b_2 = 1/2.
\end{eqnarray} 
Since the parameters $b_j$ are not integral numbers, the functions $\tilde{R}_j$ take the form \cite{Abramowitz-book}--\cite{NIST-book}
\begin{equation} \label{e: radial function tilde}
 \tilde{R}_j = e^{-y/2} \left( C_{I j} \,\, \chypergf (a_j,b_j; y) +  C_{II j} \,\, y^{1-b_j} \chypergf (a_j-b_j+1,2-b_j; y)  \right),
\end{equation} 
where $\chypergf (a_j,b_j; y)$ denotes the confluent hypergeometric function and $C_{I j}$, $ C_{II j}$ are constants. 

From the values (\ref{e: values parameters a b}) it is convenient to note that $b_1 = b_2= 1/2$ and therefore making the change of variable $y = z^2$ we can transform Eqs.\  (\ref{e: R radial y coordinate}) into the Hermite differential equations \cite{Abramowitz-book}--\cite{NIST-book}
\begin{equation} \label{e: Hermite form}
 \frac{\dd^2 \hat{R}_j}{\dd z^2} - 2 z  \frac{\dd \hat{R}_j}{\dd z} + \lambda_j \hat{R}_j = 0,
\end{equation} 
with $\lambda_j = - (1- \epsilon + 2i m^2 / \omega)$. Thus using the linearly independent solutions of Eqs.\  (\ref{e: Hermite form}) we can expand the functions $\hat{R}_j$, but to write the solutions of the differential equations (\ref{e: R radial y coordinate}) we prefer to use confluent hypergeometric functions. Furthermore, to increase the readability, in the expressions that follow we replace the quantities $b_1$ and  $b_2 $ by their numerical values, but we preserve the parameters $a_1$ and $a_2$.  

We point out that the four constants $C_{I j}$ and $ C_{II j}$ are not independent since Eqs.\ (\ref{e: coupled tilde R variable y}) impose the following restrictions:

a) If we choose the function $\tilde{R}_1$ as
\begin{equation}
 \tilde{R}_1 = C_{I 1}\, e^{-y/2}   \,\, \chypergf (a_1,1/2; y) ,
\end{equation} 
then the function $\tilde{R}_2$ must take the form 
\begin{equation}
 \tilde{R}_2 = C_{I 1}  \frac{2 (2 \omega)^{1/2} i^{1/2} a_1}{ m}  e^{-y/2} y^{1/2} \chypergf (a_2+1/2,3/2; y),
\end{equation} 
that is, the constants $C_{I 1}$ and $ C_{II 2}$ fulfill 
\begin{equation} \label{e: relation constant 1}
 C_{II 2} = C_{I 1} \frac{2 (2 \omega)^{1/2} i^{1/2} a_1}{ m}.
\end{equation} 

b) When we select the function $\tilde{R}_1 $ as
\begin{equation}
 \tilde{R}_1 = C_{I 2}\,  e^{-y/2}  y^{1/2} \chypergf (a_1+1/2,3/2; y),
\end{equation} 
then the function $\tilde{R}_2$ must be of the form
\begin{equation}
 \tilde{R}_2 = C_{I 2}  \frac{(2 \omega)^{1/2} i^{1/2} }{2 m}  e^{-y/2} \chypergf (a_2,1/2; y). 
\end{equation} 
Hence the constants $C_{I 2}$ and $ C_{II 1}$ satisfy 
\begin{equation} \label{e: relation constant 2}
 C_{II 1} = C_{I 2} \frac{(2 \omega)^{1/2} i^{1/2} }{2 m} .
\end{equation} 

From these results and considering the definition (\ref{e: R tilde definition}) we get that the linearly independent solutions $Z_\pm$ to the Schr\"odinger equations of the CES potentials (\ref{e: ces potentials}) are equal to
\begin{eqnarray} \label{e: zeta one}
 Z_\pm^I &=&  C_{I1}\, \textrm{e}^{-i \pi/4}  \textrm{e}^{-y/2} \left[ \chypergf (a_1,1/2; y) \right.  \nonumber \\  
 & &\left. \pm \frac{2 (2 \omega)^{1/2} i^{3/2} a_1}{m}  y^{1/2} \chypergf (a_2+1/2,3/2; y) \right],
\end{eqnarray}
and
\begin{eqnarray} \label{e: zeta two}
 Z_\pm^{II} &=&  C_{I 2}\, \textrm{e}^{-i \pi/4}  \textrm{e}^{-y/2} \left[ y^{1/2} \chypergf (a_1+1/2,3/2; y)  \right.  \nonumber \\  
 & &\left. \pm \frac{(2 \omega)^{1/2} i^{3/2} }{2 m}  \chypergf (a_2,1/2; y) \right] .
\end{eqnarray}

Using that for the linearly independent solutions of the confluent hypergeometric equation (\ref{e: confluent hypergeometric equation}) its Wronskian is equal to \cite{NIST-book}, \cite{Erdelyi-I-book}
\begin{equation}
 \mathcal{W}_z [\chypergf (a,b; z), z^{1-b} \chypergf (a-b+1,2-b; z) ] = (1-b) z^{-b} \textrm{e}^z,
\end{equation} 
we find that the Wronskian of the solutions (\ref{e: zeta one}) and (\ref{e: zeta two}) is (for $ C_{I1} = C_{I 2} = 1$)
\begin{equation}
 \mathcal{W}_x [Z_\pm^I,Z_\pm^{II}] = \mp \frac{\omega \, i^{3/2}}{m} (2 \omega)^{1/2} ,
\end{equation} 
which is a constant, as expected.

Moreover considering Eqs.\ (\ref{e: R tilde definition}),  (\ref{e: coupled tilde R}), and (\ref{e: r tilde differential equations}) we obtain that the functions $Z_{\pm}$ are solutions of the equations
\begin{equation}
  \frac{\dd^2 Z_{\pm}}{\dd y^2} - \frac{1}{4} Z_{\pm} - \frac{i m^2}{2 \omega y} Z_{\pm} \pm \frac{m}{2(2 \omega)^{1/2} i^{3/2} y^{3/2}} Z_{\pm} = 0. 
\end{equation} 
In a straightforward way we can verify that these are the Schr\"odinger equations of the partner potentials $V_\pm$ written using the variable $y$ defined in the formula (\ref{e: y definition}), and hence we find that the functions $\tilde{R}_j$ yield the solutions to the Schr\"odinger equations of the partner potentials (\ref{e: ces potentials}).

\section{Discussion}
\label{s: Discussion}

We have not been able to transform the linearly independent solutions (\ref{e: zeta one}), (\ref{e: zeta two}) and write them in terms of one confluent hypergeometric function, but we think that this fact must be considered in detail. Thus we find in explicit form the partner potentials (\ref{e: ces potentials}) whose linearly independent solutions include the sum of two confluent hypergeometric functions  (see  (\ref{e: R tilde definition}), (\ref{e: zeta one}), and (\ref{e: zeta two})). In Ref.\  \cite{Cooper:1986tz} some potentials with this property are studied, but they are given in implicit form. Hence the partner potentials (\ref{e: ces potentials}) expand the previously known examples of potentials whose linearly independent solutions include a sum of (confluent) hypergeometric type functions \cite{Cooper:1986tz}. 	

Furthermore the  CES potentials $V_\pm$ are multiplicative shape invariant \cite{Cooper:1994eh}, \cite{Cooper-book}, \cite{Gendenshtein}. From their expressions (\ref{e: ces potentials}) we find that they satisfy $ V_+(x,m) = V_-(x,-m)$. When we compare with the definition of shape invariance for the partner potentials \cite{Gendenshtein}
\begin{equation}
  V_+(x,\alpha_0) = V_-(x,\alpha_1) + R (\alpha_0) ,
\end{equation}
where $\alpha_0$, $\alpha_1 = f(\alpha_0)$ are parameters and $R (\alpha_0)$ is a function of $\alpha_0$, 
we find that the partner potentials $V_\pm$ fulfill $R(\alpha_0)=0$, $\alpha_0 = m$, $\alpha_1=-m$, and hence, $\alpha_1=-\alpha_0 $. We notice that  the previously published multiplicative shape invariant potentials only are known in series form \cite{Cooper:1994eh}, \cite{Cooper-book}, in contrast we find the potentials (\ref{e: ces potentials}) in closed form.

In Ref.\  \cite{Aleixo-Balantekin} is studied the concept of shape invariance with reflection transformations. For the analyzed potentials they consider transformations that include reflections of the coordinates and translations of the parameters. The relationship $\alpha_1=-\alpha_0 $ between the parameters of the partner potentials (\ref{e: ces potentials}) reminds us a reflection, but this reflection is on the parameters of the potentials and not on the coordinates, as considered in Ref.\  \cite{Aleixo-Balantekin}.

Owing to the CES potentials (\ref{e: ces potentials}) go to zero as $x \to + \infty$,  in this limit  $Z_\pm$ behave as
\begin{equation}
 Z_\mp \approx A_\mp \sin \left( \omega x + \delta_0^\mp (\omega) \right),
\end{equation}
where  $\delta_0^\pm (\omega) $  denote the phase shifts of $V_\pm$ ($ A_\mp $ are constants). Using the known relationship between the phase shifts of the partner potentials \cite{Cooper-book}, \cite{Gangopadhyaya book}
\begin{equation}
 \textrm{e}^{2 i \delta_0^- (\omega)}  = \frac{W_+ - i \omega}{W_+ + i \omega} \textrm{e}^{2 i \delta_0^+(\omega)},
\end{equation} 
where $W_+$ is the limit of the superpotential $W$ as $x \to + \infty$ (see (\ref{e: limits potentials superpotential})), we get that the phase shifts of the CES potentials (\ref{e: ces potentials}) fulfill
\begin{equation}
 \delta_0^+(\omega) = \delta_0^-(\omega) + \left(n + \frac{1}{2}\right) \pi,
\end{equation} 
where $n$ is an integer.

From our previous results we think that a research problem is the search, in explicit form, of potentials for which each linearly independent  solution of their Schr\"odinger equations includes a sum of three (or four, five, \dots) (confluent) hypergeometric functions. For the case of three (confluent) hypergeometric functions we expect that the first linearly independent solution takes the form
\begin{equation}
 Z^I = f_1 (x) F_1 (x) + f_2 (x) F_2 (x) + f_3 (x) F_3 (x) ,
\end{equation} 
where $f_k$ are functions of $x$ and $F_k$ denotes the (confluent) hypergeometric functions (a similar expression is valid for the second linearly independent solution).\footnote{See the expressions (\ref{e: zeta one}) and (\ref{e: zeta two}) for linearly independent solutions of the Schr\"odinger equation that include two confluent hypergeometric functions.} Notice that in Ref.\  \cite{Cooper:1986tz} are studied some potentials with this property, but they are given in implicit form. If we find these potentials, they can be used as a basis to generate new exactly solvable potentials by means of first or higher order supersymmetric quantum mechanics \cite{Cooper:1994eh}--\cite{Gangopadhyaya book}.

\section{Acknowledgments}


We thank the support by CONACYT M\'exico, SNI M\'exico, EDI-IPN, COFAA-IPN, and Research Projects IPN SIP-20150707 and IPN SIP-20151031. Also, we thank the useful suggestions of the anonymous Referees.



\end{document}